\documentclass[sn-mathphys-ay]{sn-jnl}

\usepackage{graphicx}%
\usepackage{multirow}%
\usepackage{amsmath,amssymb,amsfonts}%
\usepackage{amsthm}%
\usepackage{mathrsfs}%
\usepackage[title]{appendix}%
\usepackage{xcolor}%
\usepackage{textcomp}%
\usepackage{manyfoot}%
\usepackage{booktabs}%
\usepackage{algorithm}%
\usepackage{algorithmicx}%
\usepackage{algpseudocode}%
\usepackage{listings}%
\usepackage{lineno}

\theoremstyle{thmstyleone}%
\theoremstyle{thmstyletwo}%

\theoremstyle{thmstylethree}%

\begin{document}

\title[Evaluation of the Real-time El~Ni\~no Forecasts  by the  Climate Network Approach
between 2011 and Present]{Evaluation of the Real-time El~Ni\~no Forecasts  by the  Climate Network Approach
between 2011 and Present}


\author[1]{\fnm{Armin} \sur{Bunde}}

\author*[2]{\fnm{Josef} \sur{Ludescher}}\email{josef.ludescher@pik-potsdam.de}

\author[2,3]{\fnm{Hans Joachim} \sur{Schellnhuber}}

\affil[1]{\orgdiv{Institute for Theoretical Physics}, \orgname{Justus Liebig University Giessen}, \orgaddress{\postcode{35392}  \city{Giessen},  \country{Germany}}}

\affil[2]{\orgname{Potsdam Institute for Climate Impact Research (PIK), Member of the Leibniz Association}, \orgaddress{\postcode{14412} \city{Potsdam}, \country{Germany}}}

\affil[3]{\orgname{International Institute for Applied Systems Analysis (IIASA)}, \orgaddress{\postcode{2361} \city{Laxenburg},  \country{Austria}}}

\abstract{El Ni\~no episodes are part of the El~Ni\~no-Southern Oscillation (ENSO), which is the strongest driver of interannual climate variability, and can trigger extreme weather events and disasters
in various parts of the globe. Previously we have described a network approach that allows to forecast  El~Ni\~no events about 1 year ahead.
Here we evaluate
the real-time forecasts  of this   approach  between 2011 and 2022.
We find that the approach correctly predicted (in 2013 and  2017) the onset of both El~Ni\~no   periods (2014-2016 and  2018-2019) and generated only 1 false alarm in 2019.
In  June 2022, the approach correctly forecasted the onset of an  El~Ni\~no event in 2023.
We show how to determine the $p$-value of the 12 real-time forecasts between 2011 and 2022
and find $p\cong 0.005$, this way strongly rejecting the null hypothesis that the same quality of the forecast can be obtained by random guessing.
We also discuss  how  the algorithm can be further improved by reducing the number of false alarms in the network model forecast.
When combined with other statistical methods, a more detailed forecast, including the magnitude of the event and its type, can be obtained. For 2024, the method indicates the absence of a new El~Ni\~no event.}




\maketitle

\section{Introduction}\label{sec1}

The El~Ni\~no-Southern Oscillation (ENSO)
(\citealt{ Dijkstra2005, Clarke08, Sarachik10, Wang2017,Timmermann2018, McPhadden2020})
can be considered as a  quasi-oscillation of the Pacific ocean-atmosphere system, consisting of    irregular warm (``El~Ni\~no'') and cold (``La Ni\~na'') deviations from the long-term mean.
Strong El~Ni\~no episodes can lead to extreme weather events  (like extreme rainfall and droughts) in various parts of the globe (\citealt{Davis2001,Wen2002,Kovats03,Donnelly07,Corral10,McPhadden2020}).
To mitigate at least some of the adverse societal and economic impacts,
early forecasts of El Ni\~no events are thus highly desirable.

To forecast El~Ni\~no events, many state-of-the-art coupled climate models, as well as a variety of statistical approaches
(\citealt{Cane86,Penland1995,Tziperman97,Fedorov03,Galanti03,Kirtman03,Chen04,Palmer2004,Luo08,Chen08,Chekroun11,Saha2014, Chapman2015, Feng2016, Lu2016, Rodriguez2016, Meng2018,Nootboom2018, Ham2019,DeCastro2020,Petersik2020,Hassanibesheli2022}),
have been suggested, and monthly updated overviews of the latest operational forecasts (consisting of 17 dynamical and 9 statistical methods) are available from the International Research Institute for Climate and Society  (\citealt{IRI_forecast_current}). While these forecasts are quite successful at shorter lead times,
they have  limited anticipation power at larger lead times.
In particular, they generally fail to overcome the so-called ``spring barrier'' (see, e.g., \citealt{Webster1995,Goddard2001}), which shortens their typical warning time to around 6 months (\citealt{Barnston2012, McPhadden2020})  (see also the discussion in \citealt{Tippett}).

In 2012, an alternative forecasting approach (\citealt{Ludescher2012,Ludescher2013})  (see also \citealt{Ludescher2014}) has been suggested, which is  based on complex-networks analysis   (\citealt{Tsonis2006,Yamasaki2008,Donges2009,Gozolchiani2011,Dijkstra2019,Fan2021,Ludescher2021}). The approach analyses the strength of the cooperativity represented by the mean link strength $S(t)$ in a Pacific climate network, and gives an alarm when $S$ crosses a fixed threshold, predicting a new  El~Ni\~no episode  to come in the following year.  The optimal threshold $\Theta$  was determined in a learning period between 1950 and 1980. In the period between 1981 and 2011, this threshold $\Theta$ was used to hindcast the presence  (alarm) or absence (no alarm) of a new  El~Ni\~no   event in the following  year. After the threshold is fixed, there is no free parameter in the approach.

The procedure to split the known data (at that time between 1950 and 2011) into a learning phase and a hindcasting phase is necessary for statistical forecasting methods and aims to reduce the risk of an overfitting to spurious precursors.
But the mere fact that each algorithm,
when being developed, can only make ``predictions" of events that have already occurred automatically introduces a certain ``publication" bias, because only those algorithms that are successful in both the learning and hindcasting phase will be considered and published.

The true test for statistical forecasts are  real-time forecasts.  For the climate network approach, the period of real-time forecasts started in 2011. Here we evaluate the real-time forecasts of the network approach. First,
in Section 2, we describe how El~Ni\~no-events are classified by the  Oceanic Ni\~no Index (ONI) and list the ONI values between 2011 and present.
Next, in Section 3, we briefly describe the climate network approach.  In Sections 4 and 5,  we analyse its  real-time forecasts between 2011 and present  and determine the statistical  significance
of the forecast. In Section 6, we describe an improvement of the algorithm, which is based on the false alarm statistics.

\section{Data}
The ENSO phenomenon  is  quantified by the   Oceanic Ni\~no Index (ONI), which is
defined as the three-month running-mean  sea surface temperature (SST) anomalies in the Ni\~no3.4 region (see Fig. 1)
and is a principal measure for monitoring, assessing, and predicting ENSO.

An El~Ni\~no-episode is said to occur when the index is at least 0.5\textdegree C above the average for a period of at least 5 months. Table 1 shows the ONI between 2012 and present, as communicated by the  National Oceanic and Atmospheric Administration (NOAA) (\citealt{ONItable}).
The El~Ni\~no periods are in boldface. The table shows that  there were no El~Ni\~no periods in 2012, 2013, 2017, 2020, 2021, and 2022. In May 2023, an El~Ni\~no started and is still ongoing at the time of writing.

\begin{table}[ht]

\renewcommand\arraystretch{1.12}
\renewcommand\tabcolsep{3pt}

\centering
\begin{tabular}{ | c | c c c c c c c c c c c c |}
\hline
\hline
\rule[0.1mm]{0mm}{0.1mm}
Year & DJF & JFM & FMA & MAM & AMJ & MJJ & JJA & JAS & ASO & SON & OND & NDJ \\[0.5ex]
\hline

2012 & -0.9 & -0.7 & -0.6& -0.5& -0.3 & 0.0 & 0.2 & 0.4 & 0.4 & 0.3 & 0.1 & -0.2 \\
2013 & -0.4 & -0.4 & -0.3 & -0.3 & -0.4 & -0.4 & -0.4 & -0.3 & -0.3 & -0.2 & -0.2 & -0.3 \\
2014 & -0.4 & -0.5 & -0.3 & 0.0 & 0.2 & 0.2 & 0.0 & 0.1 & 0.2 & \bf{0.5} & \bf{0.6} & \bf{0.7} \\
2015 & \bf{0.5} & \bf{0.5} & \bf{0.5} & \bf{0.7} & \bf{0.9} & \bf{1.2} & \bf{1.5} & \bf{1.9} & \bf{2.2} & \bf{2.4} & \bf{2.6} & \bf{2.6} \\
2016 & \bf{2.5} & \bf{2.1} & \bf{1.6} & \bf{0.9} & {0.4} & -0.1 & -0.4 & -0.5 & -0.6 & -0.7 & -0.7 & -0.6 \\
2017 & -0.3 & -0.2 & 0.1 & 0.2 & 0.3 & 0.3 & 0.1 & -0.1 &-0.4 &-0.7 &-0.8 &-1.0 \\
2018 & -0.9 & -0.9& -0.7& -0.5& -0.2 & 0.0 & 0.1 & 0.2
 & \bf{0.5} & \bf{0.8} & \bf{0.9} & \bf{0.8}\\
2019 & \bf{0.7}& \bf{0.7}& \bf{0.7}& \bf{0.7}& \bf{0.5}& \bf{0.5}& 0.3 & 0.1 & 0.2 & 0.3 & 0.5 & 0.5 \\

2020 & 0.5 & 0.5 & 0.4 & 0.2 & -0.1 & -0.3 & -0.4 & -0.6 & -0.9 & -1.2 & -1.3 & -1.2 \\
2021 & -1.0 & -0.9 & -0.8& -0.7& -0.5 & -0.4 & -0.4 & -0.5 & -0.7 & -0.8 & -1.0 & -1.0 \\
2022 & -1.0 & -0.9 & -1.0& -1.1& -1.0 & -0.9 & -0.8 & -0.9 & -1.0 & -1.0 & -0.9 & -0.8 \\

2023 & -0.7 & -0.4 & -0.1 & 0.2 & \bf{0.5} & \bf{0.8} & \bf{1.1} & \bf{1.3} & \bf{1.6} & \bf{1.8} & \bf{1.9} & \bf{2.0} \\
\hline

\end{tabular}
\caption{Oceanic El~Ni\~no Index (ONI) 2012 - present (from \citealt{ONItable})}
\label{table1}
\end{table}

\section{ The climate network approach}

The structure of the climate network considered here is shown in Fig. 1.
The network is based on a combination of the networks introduced by (\citealt{Yamasaki2008}) and (\citealt{Gozolchiani2011}), who studied cooperative phenomena during El~Ni\~no events.
The nodes of the network consist of 14 grid points in the ``El~Ni\~no basin'' (red circles) (\citealt{Gozolchiani2011})  (which roughly covers  the Ni\~no1, Ni\~no2, Ni\~no3, and Ni\~no3.4 regions), and 193 grid points outside this domain (blue  circles) (\citealt{Yamasaki2008}).

\begin{figure}[h!]
\begin{center}
\includegraphics[width=7cm]{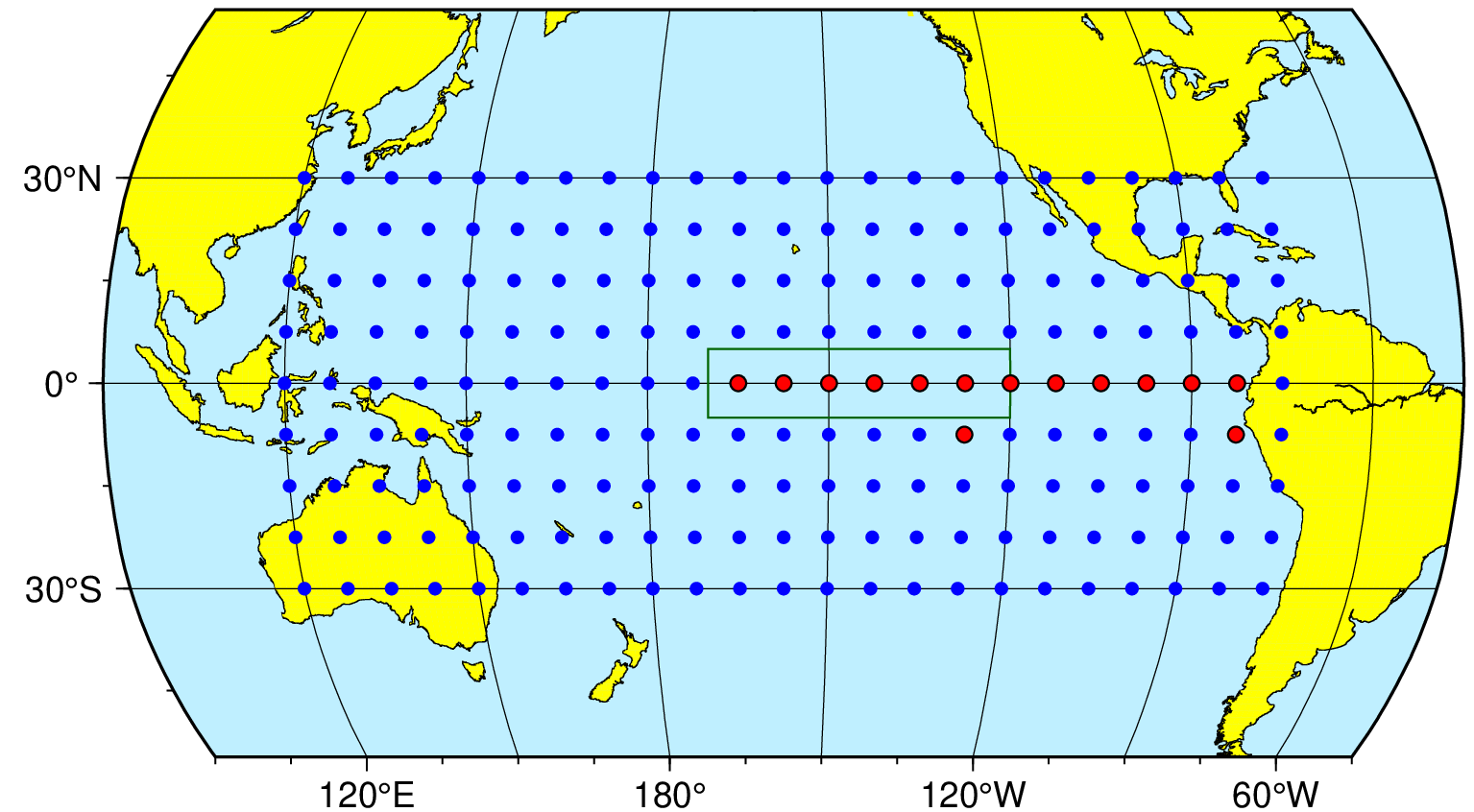}
\caption{The  structure of the climate network. Each of the  14 grid
points  in the ``El~Ni\~no basin''  (red circles) is linked to each of the 193 grid points outside this
domain ({blue circles}). The green rectangle denotes the Ni\~no3.4 region.}
\label{fig1}
\end{center}
\end{figure}

The green rectangle denotes the Ni\~no3.4 region where the ONI is calculated.
The grid points are the nodes of the climate network and are characterized by their surface air temperature (SAT) anomaly. The SAT data are obtained from the NCEP Reanalysis 1 dataset (\citealt{Kalnay1996, reanalysis2}).

Each node inside the El~Ni\~no basin is linked to each node outside the basin.
The link strength between two nodes (i.e., the strength of the teleconnections between them) at a given time $t$ is determined from the values of their time-lagged cross-correlation (see Appendix A) for which we consider time lags between 0 and 200 days.
For each pair of nodes $i$ and $j$, we determine, for the given time $t$, the maximum, the mean, and the standard deviation around the mean of the absolute value of the cross-correlation function, and define the
link strength $S_{ij }(t)$ as the difference between the maximum and the mean value, divided by the standard deviation. Accordingly,
$S_{ij }(t)$ describes the link strength relative to the underlying background noise (signal-to-noise ratio).
By averaging over all individual links in the network at a given instant $t$, one obtains the mean link strength $S(t)$, which is the crucial entity in the climate network approach (for details, see (\citealt{Gozolchiani2011,Ludescher2013}) and Appendix A). The variation of $S(t)$ with time $t$ can be considered as a measure of the way the cooperativity between the equatorial ``El~Ni\~no  basin" and the rest of the tropical and subtropical Pacific region changes with time $t$.
$S(t)$ has a remarkable property: it typically decays during an  El~Ni\~no   event (\citealt{Ludescher2013}) and rises in the year before an event starts.
This rise of $S(t)$  can be  used as a precursor for the event (\citealt{Ludescher2013, Ludescher2014}).

The optimized algorithm involves an empirical decision threshold $\Theta$. Whenever $S$ crosses $\Theta$ from below while the most recent  ONI
(see Data section) is below 0.5\textdegree C, the algorithm sounds an alarm and predicts  the start of a new El~Ni\~no episode in the following year. Otherwise, it predicts the absence of a new El~Ni\~no event.

\begin{figure}[h!]
\begin{center}
\includegraphics[width=\linewidth]{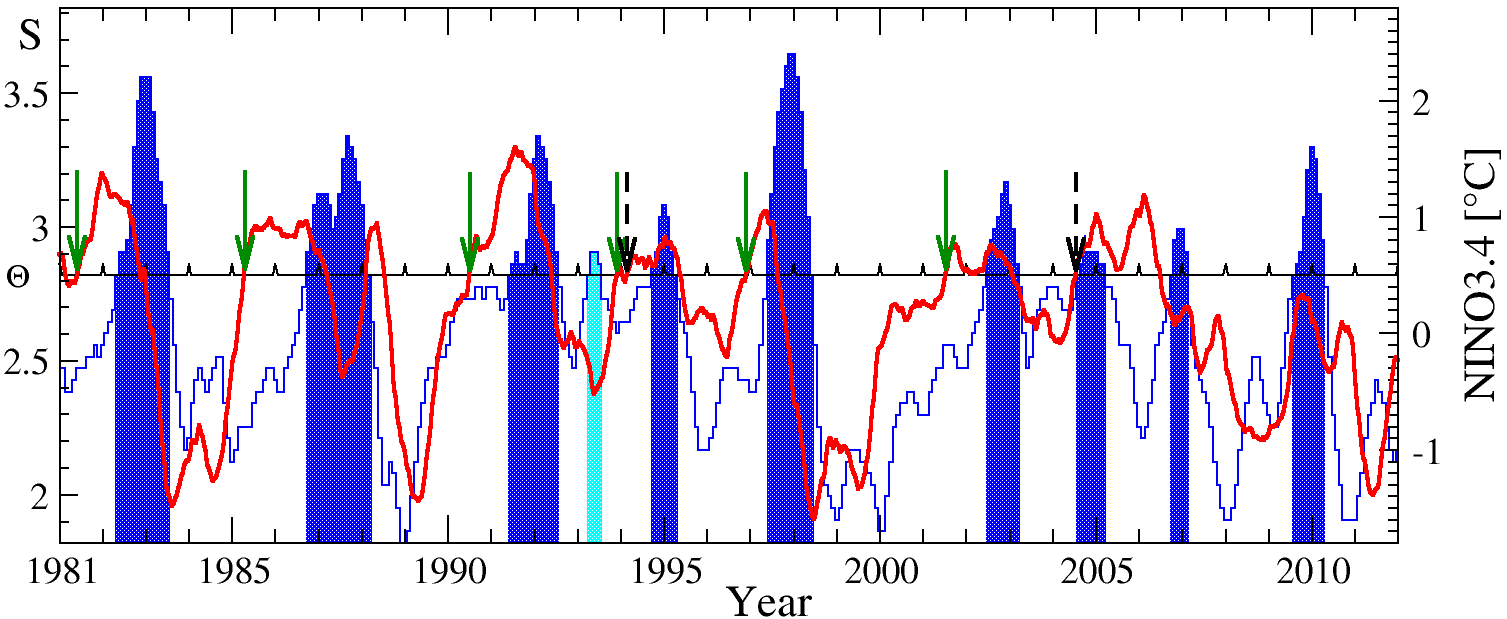}
\caption{ The forecasting scheme.  We compare the average link strength $S(t)$
in the climate network (red curve) with the decision threshold $\Theta=2.82$ (horizontal line)
and the ONI (right scale), between January  1981 and December 2011.
When the link strength crosses the threshold from below and the last available ONI is below 0.5\textdegree C,
 we give an alarm and predict that a new El~Ni\~no episode will start in the following calendar year.
Periods with an ONI  greater or equal  0.5\textdegree C are displayed in blue. The El~Ni\~no episodes (when the ONI is greater or equal 0.5\textdegree C for at least 5 months) are displayed in dark blue.
Correct predictions are marked by green arrows and false alarms by dashed arrows. Note that the early false alarms in February 1994 and July 2004 are followed by at least one
ONI  value equal or above 0.5\textdegree C in the same year.
}
\label{fig2}
\end{center}
\end{figure}
In the learning phase between 1950 and 1980,
all thresholds above the temporal mean of $S(t)$ were considered
and the optimal ones, i.e., those ones that lead to the best predictions in the learning phase, were  determined.
$\Theta$-values between $2.815$ and $2.834$ lead to the best performance (\citealt{Ludescher2013}), with a false alarm rate of 1/20.

In the hindcasting phase (1981-2011)  (see Fig. 2, where $\Theta=2.82$), the performance of these thresholds was tested; thresholds between 2.815 and 2.826 gave the best results. Figure 2 shows that the alarms were correct in 75\% and the non-alarms in 86\% of
all cases. For $\Theta$-values between $2.827$ and $2.834$, the performance was only slightly weaker.

\section{ Real-time forecasts between 2011 and  present}
Figure 3 shows the forecasts of the network approach between 2011 and 2022.
\begin{figure}[h!]
\begin{center}
\includegraphics[width=\linewidth]{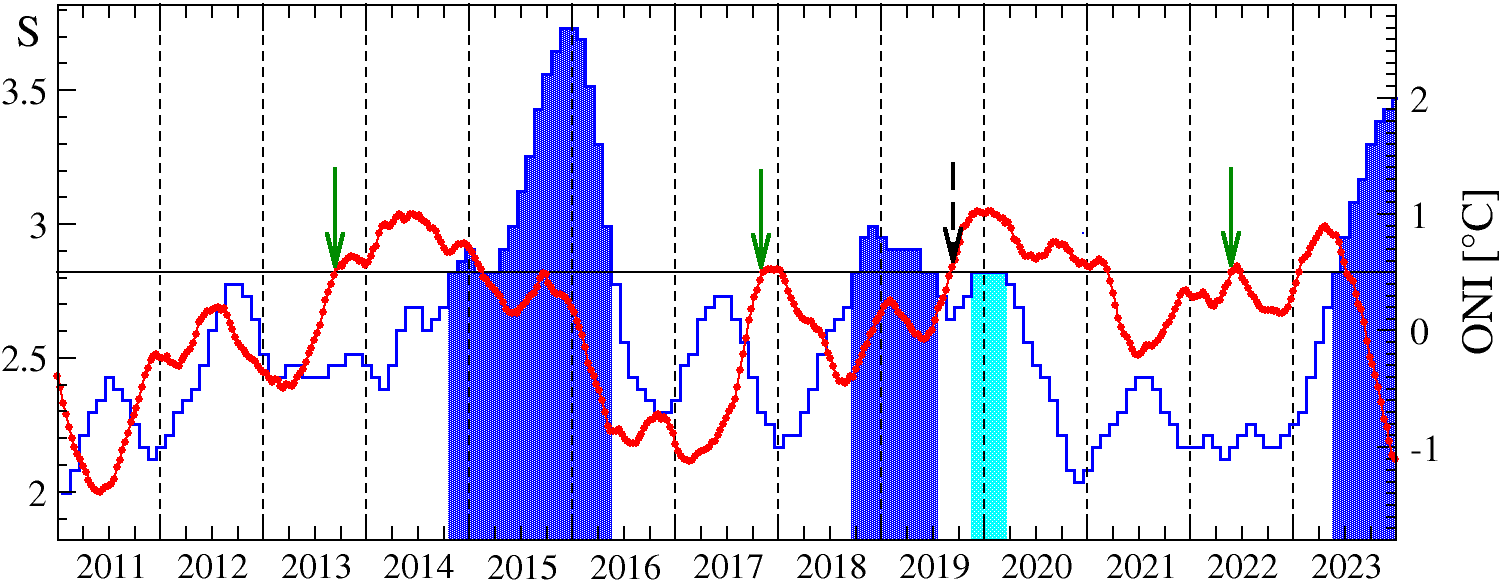}
\caption{
The real-time forecasts. Same as Fig. 2, but for the period between January 2011 and December 2023. As in Fig. 2, the false alarm (in 2019) is
followed by at least one ONI  value equal to or above 0.5\textdegree C in the same year. Only alarms until 2022, where the outcome is known, are marked by arrows.}
\label{fig3}
\end{center}
\end{figure}
In 4 years (2013, 2017,  2019, and 2022) the algorithm predicted the onset of a new El~Ni\~no event in the following calendar year. Only the alarm of 2019
was a false alarm. The present El~Ni\~no started in May 2023, so the alarm given in June 2022 was also correct.

In   8 years (2011, 2012, 2014, 2015,  2016, 2018, 2020, 2021)  the approach did not give an alarm and thus correctly predicted the absence of a new  El~Ni\~no   in the following year.
This is true also for 2014, since in 2015 no {\it new}  El~Ni\~no  episode, separated from the foregoing one by at least one ONI value below 0.5, started.
Also these forecasts of the {\it absence} of a new  El~Ni\~no event are far from being trivial as a comparison with the official forecasts by the International Research Institute for Climate and Society  (\citealt{IRI_forecast})
shows:

(i) While the climate network approach already in December 2011 indicated the absence of a new El~Ni\~no in 2012,
the CPC/IRI consensus probabilistic ENSO forecast provided in August and September 2012 75 and 65 percent probability, respectively, for the presence of El Ni\~no conditions in December 2012 (NDJ).

(ii) In spring 2017, most dynamical and statistical models falsely predicted an event
in 2017. For instance, the vast majority of the ensemble members  of the North American Multimodel forecasted, in April 2017,
positive anomalies, while the actual SSTA turned out to be negative (\citealt{Tippett}).

Indeed, according to (\citealt{Tippett}), climate models tend to predict warming when initialized after observed warming conditions and cooling when initialized after observed cooling conditions, and thus failed to capture the correct direction of ENSO evolution in half of the 8 springs between 2011 and 2018.

Next, we turn to the question whether the real-time  forecasts of the climate network approach  are statistically significant, i.e., whether the same performance can be obtained by random guessing or not.

\section{Statistical significance of the forecasts}
For obtaining the statistical  significance of a given configuration $K_0$ containing  $N$ forecasts with $n_c$ correct alarms and $n_f$ false alarms, one has to determine the probability $w_0$ that a configuration with the same number $n_c$ of correct alarms and the same number $n_f$ of false alarms can be obtained by randomly guessing. In addition, one has to consider all configurations $K_1, K_2, \dots, K_m$ with a better or equal quality of forecast and determine the corresponding probabilities $w_1, w_2, \dots, w_m$.  Then the  probability $p$ that by randomly guessing the same or better forecasts can be made is given by

\begin{equation}
p=\sum_{i=0}^m w_i,
\end{equation}
$p$ is called the $p$-value. In our case, the null hypothesis is that the given forecast configuration can be obtained by randomly  guessing with the climatological El Ni\~no onset probability.  When $p$ is below 0.05, the null hypothesis is rejected and the forecasts are called statistically significant at a 0.05 level; when $p$ is below 0.01, the forecasts are called highly significant.
For determining the probabilities $w_i$, we first need to determine the occurrence probability  $q$ of the onset of  El~Ni\~no   episodes. In the 43 years between January 1981 and 2023, 12 El~Ni\~no   episodes started, so the occurrence probability is $q=12/43\cong 0.279$.
First we focus  on the occurrence of new El~Ni\~no   episodes in the period between  January 2012 and December 2023. Denoting  years where a new  event started by $+$ and years where no new event started by $-$, the observed configuration of years with and without new El~Ni\~no events is

 \begin{equation}
(-,-,+,-,-,-,+,-,-,-,-,+),
\end{equation}
where the most left symbol refers to 2012 and the right-most symbol to 2023 where a new  El~Ni\~no   episode started in May.
For the period between 2012 and 2023, the network approach predicted the configuration

\begin{equation}
(-,-,+,-,-,-,+,-,+,-,-,+),
\end{equation}
which differs from the observed configuration only in the year 2020 (+ instead -), where a new  event  was falsely predicted to come.

There are 9 possible configurations where one of the $ - $ signs in the observed configuration (2) is changed into a $+$ sign, and all have the same quality of forecast.
Accordingly, the probability of randomly guessing one of these 9 configurations is
$w_0=9 q^4(1-q)^8$.

There is only one better forecast possible: the probability $w_1$ of randomly guessing the observed configuration (2) is
$w_1=q^3(1-q)^9$. Accordingly, the $p$-value of the real-time forecasts is

\begin{equation}
p=9q^4(1-q)^8 + q^3(1-q)^9.
\end{equation}
This yields, with $q=12/43$,
\begin{equation}
p\cong 5.1 \times10^{-3},  \\\  {\rm period \\\ 2011-present},
\end{equation}
which is well below the  high-significance  threshold $p=10^{-2}$.

When we consider both the hindcasting and forecasting period (January 1981 - December 2023) the $p$-value is obtained in exactly the same way, but there are more configurations to be considered. In the 43 years between 1981 and 2023, 12 new  El~Ni\~no   episodes started.  In the 42  target years between 1982 and 2023, the network algorithm correctly forecasted  9 of these events and gave 3 false alarms.
Accordingly, the hit rate $\alpha_+$ defined as the number of correct alarms $n_c$ divided by the number of  events, is 9/12, while the false alarm rate, defined as the number of false alarms $n_f$ divided by the number of non-events, is 1/10. Thus the rate $\alpha_-$ of correctly predicted non-events is $(30-3)/30=9/10$.
Both numbers, $\alpha_+$ and $\alpha_-$  quantify the performance of the algorithm.
The probability of randomly guessing a configuration with $n_c$ correct events and $n_f$ false events is given by

\begin{equation}
w=\binom{12}{n_c}\binom{30}{n_f}q^{n_c+n_f}(1-q)^{42-n_c-n_f}.
\end{equation}
The binomial coefficients describe the number of ways $n_c$ events can be chosen out of 12 events and $n_f$ false events out of 30 non-events;  $q=12/43$ as above.

We need to determine $w$ for all configurations with a similar or better predictive power. A natural measure for the predictive power is
$P={(\alpha_+ + \alpha_-) -1}$, which is 1 when the forecast is perfect and 0 when the forecast is purely random.
Here, $P=(3/4+ 9/10)-1 = 0.65$.

Accordingly, for estimating the $p$-value of our forecast, we take into account all configurations with  a higher or equal predictive power, i.e.,  ($n_c=8, n_f=0$),  ($n_c=9, n_f=0,1,2, 3$), ($n_c=10, n_f=0,1,\dots,5$), ($n_c=11, n_f=0,1,\dots,8$),
and ($n_c=12, n_f=0,1,\dots,10$). For each of these  combinations of $(n_c, n_f)$, we determine $w$ from (6)
and sum up (1) the obtained probabilities.
The result is

\begin{equation}
p\cong 3.0  \times10^{-5},  \\\  {\rm period \\\ 1981-present}.
\end{equation}

\section{Further improvement of the algorithm based on the false alarm characteristics }
Figures 2 and 3 show that all false alarms in the hindcasting and forecasting period (1994, 2004, and 2019)
are followed by at least one ONI value equal or above 0.5\textdegree C in the same calendar year. This suggests that there may be only a  low chance that   an alarm is correct when the ONI does not stay below 0.5\textdegree C for the rest of the year.
Accordingly, an improved algorithm   based on this feature may consist of 2 steps. (i) In the first step, a (preliminary) alarm is given when $S$ crosses the threshold from below, indicating the  possible appearance of an El Ni\~no event in the following year. This alarm can occur at any time in a calendar year. (ii)  When the ONI stays below 0.5\textdegree C until the end of December,
this alarm is confirmed. Otherwise, the alarm is withdrawn and the absence of an El~Ni\~no onset is predicted for the following year.

When applying the improved algorithm to the period between 2011 and 2022, all forecasts turn out to be correct, resulting  in a $p$-value $p=q^3(1-q)^9\cong 1.1\times 10^{-3} $.
For the hindcasting plus forecasting period (1981-present), all 9 El~Ni\~no onset alarms are correct and there are 3 missed El~Ni\~no events resulting in $p\cong 1.4\times 10^{-6}$.

Figure 3 shows that in January 2023, the mean link strength $S(t)$ crossed the threshold $\Theta$, giving a preliminary alarm. However, since an El~Ni\~no started afterwards, the ONI did not stay below 0.5\textdegree C until December 2023. Therefore, this preliminary alarm is withdrawn, indicating the absence of an  El~Ni\~no onset in 2024.
Since there were 3 missed El~Ni\~no events between 1981 and 2023 and 30 correct predictions for the absence of a new El~Ni\~no, the probability for the absence of an El~Ni\~no onset in 2024 is 30/33 $\approx 90.9\%$.

\section {Conclusions}
In summary, we have evaluated the quality of the   real-time El~Ni\~no forecasts made by  the climate network approach.
We have shown how to determine the statistical significance of the forecasts and found that its $p$-value is well below the generally accepted high-significance level $p=0.01$, this way clearly rejecting the null hypothesis that the same performance might be obtained by simple random guessing.
We are not aware of any other method that allows, within a period of 12 years, a similar quality of real-time forecasts with a lead time of about 1 y.

The climate network approach suggests that the emergence of cooperativity between the El Ni\~no basin and the rest of the Pacific  is an important prerequisite for the development of an El Ni\~no event in the following year. We can speculate that the westerly wind bursts are more effective in initiating a large scale El Ni\~no event when the Pacific is in a cooperative state, and this would explain the success of the complex network approach. But, a detailed analysis remains for future work.

The high prediction skill of the forecast and its long lead time  should allow early mitigation methods.  One of the advantages of the network approach is that it does not contain a freely choosable fit parameter. The  underlying climate network was introduced in a different context and
independently of any El Ni\~no forecasting well before  it was used to forecast
El~Ni\~no events. Also the parameters used in the calculation of the link strengths had been fixed before (\citealt{Yamasaki2008}). The only new parameter in the algorithm, the threshold $\Theta$, was fixed in the learning phase (\citealt{Ludescher2013}). The reanalysis (NCEP) temperature data   can be  easily obtained from (\citealt{reanalysis2}).
Since also the calculation of the link strengths  is straightforward and not computationally demanding, the network approach can be easily used to obtain real-time El~Ni\~no forecasts, which is an additional advantage besides the long lead time.

The climate network-based approach discussed here forecasts the onset or absence of an El~Ni\~no event in the following calendar year with high accuracy.
The approach can be combined (\citealt{Ludescher2023a}) with additional statistical forecasting methods for the magnitude (\citealt{Meng2020}) and type (\citealt{Ludescher2023b}) of an event.
This way, the events's risk potential can be estimated much in advance, and thus, more time becomes available to plan and implement adapted mitigation measures.

So far, the climate network approach has been applied only to forecasting the onset of an El Ni\~no episode. It is an open question, how to extend it to early forecast also La Ni\~na episodes. The majority of El Ni\~no episodes, in particular the strong ones, are followed by a La Ni\~na in the consecutive year, so here, the forecast is more straightforward. But often, 2-year or even 3-year La Ni\~na episodes, like the one between 2020 and 2023, occur, and the challenge is to predict both the onset and the length of a La Ni\~na episode. We think that a combination of the climate network approach with deterministic approaches that can take advantage of ENSO's quasi-oscillatory nature may be instrumental in developing an early forecasting approach for La Ni\~na episodes.

\begin{appendices}

\section{Calculation of the mean link strength in the network approach}\label{secA1}

This Appendix follows closely (\citealt{Ludescher2014}). For the prediction of the onset of  El~Ni\~no events or non-events we use the cooperative
behavior of the atmospheric temperatures in the Pacific as a precursor. To obtain a measure for the
cooperativity, we consider the daily surface atmospheric temperature anomalies (SATA) between January 1950 and December 2023 at the grid points (''nodes'') of a Pacific network, see Fig. \ref{fig1}.

We analyse the time evolution of the teleconnections (``links'') between  the temperatures at nodes $i$ inside the  ``El~Ni\~no basin'' and
nodes $j$ outside the basin. The strengths of these links are represented by the strengths of the cross correlations between the temperature records at
these sites (\citealt{Yamasaki2008}).

The prediction algorithm (\citealt{Ludescher2013, Ludescher2014}) is as follows:

(1) At each node $k$ of the network shown in Fig. 1, the daily atmospheric temperature anomalies $T_k(t)$ (actual temperature
value minus climatological average  for each calendar day, see below) at the surface area level is determined.
For the calculation of the climatological average, the leap days have been removed.
The data have been obtained from the National Centers for Environmental Prediction/National Center
for Atmospheric Research Reanalysis I project (\citealt{Kalnay1996, reanalysis2}).

(2) For obtaining the time evolution of the strengths of the links between the nodes $i$ inside the El~Ni\~no basin and the nodes $j$
outside we compute, for each 10th day $t$ in the considered time span between January 1950 and December 2023, the time-delayed cross-correlation
function defined as
\begin{equation*}
 C_{i,j}^{(t)}(-\tau)=\frac{\langle T_i(t)T_j(t-\tau)\rangle-\langle T_i(t)\rangle\langle T_j(t-\tau)\rangle}{\sqrt{\langle(T_i(t)-\langle T_i(t)\rangle)^2 \rangle}\cdot\sqrt{\langle(T_j(t-\tau)-\langle T_j(t-\tau)\rangle)^2 \rangle}}
\end{equation*}
and
\begin{equation*}
 C_{i,j}^{(t)}(\tau)=\frac{\langle T_i(t-\tau)T_j(t)\rangle-\langle T_i(t-\tau)\rangle\langle T_j(t)\rangle}{\sqrt{\langle(T_i(t-\tau)-\langle T_i(t-\tau)\rangle)^2 \rangle}\cdot\sqrt{\langle(T_j(t)-\langle T_j(t)\rangle)^2 \rangle}}
\end{equation*}
where the brackets denote an average over the past 365 d, according to
\begin{equation*}
 \langle f(t) \rangle = \frac{1}{365} \sum_{m=0}^{364} f(t-m).
\end{equation*}
We consider time lags $\tau$ between 0 and 200 d, where a reliable estimate of the background noise level can be guaranteed.

(3) We determine, for each point in time $t$, the maximum, the
mean, and the standard deviation around the mean of the absolute value of the cross-correlation function
 $|C_{ij}^{(t)}(\tau)|$
 and define the link
strength $S_{ij}(t)$ as the difference between the maximum and the
mean value, divided by the standard deviation. Accordingly, $S_{ij}$ describes the
link strength at day t relative to the underlying background noise (signal-to-noise ratio) and
thus quantifies the dynamical teleconnections between nodes i and j.

(4) To obtain the desired mean strength $S(t)$ of the dynamical teleconnections in the climate network, we simply average over all individual link strengths.

(5) Finally, we compare $S(t)$ with a decision threshold $\Theta$. When the link strength $S(t)$ crosses the threshold from below and  the last available ONI at that time $t$ is below 0.5\textdegree C, we give an alarm and predict that an El~Ni\~no episode will start in the following calendar year.

We like to add that for the calculation of the climatological average in the learning phase, all data within this time window
have been taken into account, while in the prediction phase, only data from the past up to the prediction date have been considered.

\end{appendices}

\section*{Declarations}

\begin{itemize}
\item {\bf Funding}
JL was supported by the ''Brazil East Africa Peru India Climate Capacities (B-EPICC)`` project, which is part of  the International Climate Initiative (IKI) of the German Federal Ministry for Economic Affairs and Climate Action (BMWK) and implemented by the Federal Foreign Office (AA).

\item {\bf Competing interests}
The authors declare no competing interests.

\item {\bf Data availability}
The data sources for this study are publicly available and referenced in the text.

\end{itemize}

\end{document}